\documentclass[conference]{IEEEtran}
\IEEEoverridecommandlockouts
\usepackage{cite}
\usepackage{amsmath,amssymb,amsfonts}
\usepackage{graphicx}
\usepackage{textcomp}
\usepackage{xcolor}
\usepackage{booktabs}
\usepackage{multirow}
\usepackage{url}
\usepackage{microtype}
\usepackage{tikz}
\usepackage{hyperref}
\usetikzlibrary{positioning,arrows.meta,fit,backgrounds,matrix}
\usepackage{listings}
\begin{document}

\title{SensorWF: A FAIR Generalizable Workflow Framework for Scientific Time-Series Analysis
\thanks{This paper has been accepted for publication in the Proceedings of the ReWorDS26 Workshop, held in conjunction with the 22nd IEEE International Conference on eScience (eScience 2026). \copyright~2026 IEEE. Personal use of this material is permitted. Permission from IEEE must be obtained for all other uses, in any current or future media, including reprinting/republishing this material for advertising or promotional purposes, creating new collective works, for resale or redistribution to servers or lists, or reuse of any copyrighted component of this work in other works.}}

\author{
  \IEEEauthorblockN{Logan Luna}
  \IEEEauthorblockA{
    \textit{School of Computer Science, College of Computing}\\
    \textit{Georgia Institute of Technology}\\
    Atlanta, Georgia, USA\\
    lluna@gatech.edu
  }
  \and
  \IEEEauthorblockN{
    Joseph Rigo,
    Kellan Shew,
    Raul Alejandro Vargas-Acosta
  }
  \IEEEauthorblockA{
    \textit{Department of Electrical Engineering and Computer Science}\\
    \textit{Embry-Riddle Aeronautical University}\\
    Daytona Beach, Florida, USA\\
    \{rigoj1, shewk1\}@my.erau.edu,
    alejandro.vargas@erau.edu
  }
}

\maketitle

\begin{abstract}
Scientific sensor data is foundational across a wide range of disciplines, including spacecraft engineering, clinical medicine, and atmospheric science. In each context, pipelines are constructed to ingest raw archives, assess data quality, perform feature engineering, perform semantic annotation, and record provenance. However, these pipelines are often implemented as monolithic, domain-specific scripts with implicit assumptions and limited reusability across fields. This work introduces SensorWF, a FAIR-annotated workflow framework designed for generalizable scientific time-series analysis. The framework features a five-module reusable core (M1–M5) with typed input/output contracts that constitute the analytical backbone. A domain adapter pattern isolates all domain-specific logic within M1, enabling modules M2–M5 to operate identically across disciplines. Domain assumptions are encoded in a machine-readable component registry, facilitating integration and reuse of any sensor domain without modifying the analytical core. Domain-specific analyses are incorporated as use-case extensions without altering the core modules. The framework generates runtime PROV-O/ProvONE provenance traces, including SHA-256 checksums for all file-path entities, and emits SSN/SOSA-aligned OWL ontologies as primary outputs. To assess generalizability, SensorWF is instantiated in three distinct scientific domains: spacecraft telemetry, ambulatory ECG, and atmospheric climate. Synthetic fault injection and multi-detector anomaly detection are demonstrated as representative use-case extensions. Results indicate that a single codebase, parameterized solely through M1 adapters of approximately 170–500 lines each depending on the domain's complexity, supports comprehensive analytical pipelines across domains with varying sampling rates, channel counts, and fault taxonomies. All code, component registry, ontology artifacts, and datasets are made available as an open scientific object. Our codebase is publicly available at \href{https://purl.archive.org/sensor-wf}{https://purl.archive.org/sensor-wf}.
\end{abstract}



\section{Introduction}

Disciplines as different as spacecraft operations, clinical monitoring, and atmospheric science commonly run their sensor data through similar stages. Raw measurements are ingested from an archive, screened and corrected for artifacts, and transformed into analysis-ready feature representations before any domain question is asked of them. What differs between fields is not the shape of this pipeline but the assumptions embedded at each stage: expected sampling rates, quality thresholds, and the semantics of individual channels. Those assumptions determine how a result should be interpreted, yet they are seldom documented explicitly and almost never expressed in machine-readable form.

Despite this shared structure, pipelines are typically rebuilt from scratch in each domain as monolithic, single-purpose scripts. The cost is threefold: analytical code cannot be reused across fields, cross-domain comparisons rest on implementations that were never verified to behave alike, and post-publication auditing becomes impractical because the assumptions behind a result are recoverable only by reading the source. The eScience and open-science communities have responded by arguing that pipelines should be treated as first-class scientific objects, with provenance that is traceable, shareable, and reusable \cite{wilkinson2016fair,garijo2022fair}.

Acting on that position requires an architecture that separates the analytical stages every domain shares from the assumptions each domain supplies. This paper presents \textbf{SensorWF}, a framework for multi-domain scientific sensor time-series analysis built on exactly that separation. Its reusable core comprises five modules, ingestion (M1), quality assessment (M2), feature engineering (M3), semantic annotation (M4), and provenance export (M5), of which only M1 is domain-specific; the remaining four are configured by the adapter rather than rewritten for it. Analyses particular to a discipline attach as \emph{use-case extensions} instead of modifications to the core, with synthetic fault injection (E1) and automated anomaly detection (E2) serving as demonstrations throughout. Figure~\ref{fig:architecture} illustrates the complete architecture.

\subsection{Key Contributions}

The primary contributions of this work are as follows:

\begin{figure*}[t]
\centering
\begin{tikzpicture}[
  box/.style={draw, rounded corners=3pt, minimum width=1.65cm, minimum height=0.63cm,
              align=center, font=\small\bfseries, fill=white, inner sep=3pt},
  src/.style={draw, rounded corners=2pt, minimum width=1.25cm, minimum height=0.46cm,
              align=center, font=\scriptsize, fill=gray!9, inner sep=2pt},
  ext/.style={draw, rounded corners=3pt, minimum width=1.65cm, minimum height=0.63cm,
              align=center, font=\small\bfseries, fill=orange!9, inner sep=3pt},
  arr/.style={-{Latex[length=2.5mm]}, thick},
  darr/.style={-{Latex[length=2mm]}, thick, dashed, gray!55},
  node distance=0.5cm and 0.78cm
]
\node[src] (dsrc1) {Satellite\\\tiny SCOTTI/SATLL};
\node[src, right=0.38cm of dsrc1] (dsrc2) {ECG\\\tiny MIT-BIH};
\node[src, right=0.38cm of dsrc2] (dsrc3) {Climate\\\tiny Jena};

\node[box, below=0.55cm of dsrc2, fill=teal!13] (nm1)
  {M1\\\scriptsize Domain Adapter};
\draw[arr] (dsrc1.south) -- ++(0,-0.08) -| (nm1.north west);
\draw[arr] (dsrc2.south) -- (nm1.north);
\draw[arr] (dsrc3.south) -- ++(0,-0.08) -| (nm1.north east);

\node[box, right=0.78cm of nm1, fill=blue!7] (nm2) {M2\\\scriptsize Quality};
\node[box, right=0.78cm of nm2, fill=blue!7] (nm3) {M3\\\scriptsize Features};
\node[box, right=0.78cm of nm3, fill=blue!7] (nm4) {M4\\\scriptsize Semantic};
\node[box, right=0.78cm of nm4, fill=blue!7] (nm5) {M5\\\scriptsize Provenance};

\draw[arr] (nm1) -- (nm2);
\draw[arr] (nm2) -- (nm3);
\draw[arr] (nm3) -- (nm4);
\draw[arr] (nm4) -- (nm5);

\node[ext, below=1.05cm of nm3] (ne1) {E1\\\scriptsize Injection};
\node[ext, right=0.78cm of ne1] (ne2) {E2\\\scriptsize Detection};

\draw[arr] (nm3.south) -- ++(0,-0.26) -- (ne1.north);
\draw[arr] (ne1) -- (ne2);
\draw[darr] (ne2.north) -- ++(0,0.22) -| (nm4.south);

\node[font=\tiny, right=0.18cm of nm5, align=left, text=gray!65]
  {provenance.ttl\\kg.ttl / .owl};

\begin{scope}[on background layer]
  \node[draw=teal!65, dashed, rounded corners=5pt, fill=teal!3, inner sep=6pt,
        fit={(dsrc1)(dsrc2)(dsrc3)(nm1)},
        label={[font=\tiny\itshape,text=teal!65]left:Domain Adapters}] {};
  \node[draw=blue!55, dashed, rounded corners=5pt, fill=blue!3, inner sep=6pt,
        fit={(nm2)(nm3)(nm4)(nm5)},
        label={[font=\tiny\itshape,text=blue!65]above:Reusable Core (M2--M5)}] {};
  \node[draw=orange!65, dashed, rounded corners=5pt, fill=orange!3, inner sep=6pt,
        fit={(ne1)(ne2)},
        label={[font=\tiny\itshape,text=orange!65]below:Use-Case Extensions (opt.)}] {};
\end{scope}
\end{tikzpicture}
\caption{SensorWF architecture. The domain adapter (teal, M1) normalizes raw sensor archives from three domains into a standardized DataFrame. The reusable core (blue, M2--M5) performs quality assessment, feature engineering, semantic annotation, and provenance export identically across all domains. Optional use-case extensions (orange, E1--E2) inject synthetic faults and evaluate ML detectors; E2 results optionally enhance M4's knowledge graph (dashed arrow). Switching domains requires only a new M1 adapter ($\approx$150\,lines of Python); M2--M5 require no modification.}
\label{fig:architecture}
\end{figure*}
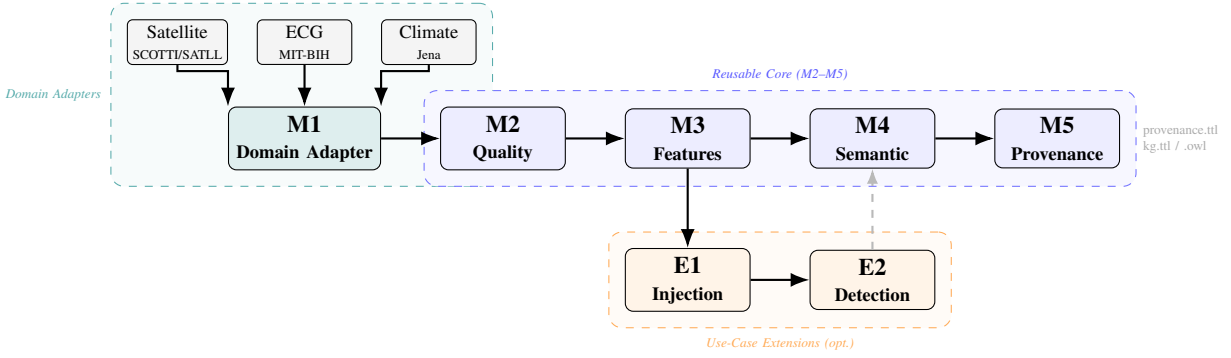

\begin{enumerate}
  \item \textbf{A generalizable five-module workflow core} (M1--M5) featuring typed input/output contracts and explicit scientific assumptions. The \emph{domain adapter} pattern confines domain-specific logic to M1, while M2--M5 operate uniformly across disciplines through adapter-driven configuration.

  \item \textbf{A machine-readable component registry} (\texttt{components.json}) that encodes domain assumptions as typed ports, configuration parameters, domain tags, and ProvONE URIs for each module. This design enables integration and reuse of any sensor domain without modifying the analytical core, thereby supporting automated composition, discovery, and extension.

  \item \textbf{Runtime provenance and semantic annotation.} SensorWF generates PROV-O/ProvONE Turtle provenance for each execution, including SHA-256 checksums for all file-path entities. Analytical evidence is mapped to domain OWL ontologies (aligned with SSN/SOSA \cite{haller2019ssn,compton2012ssn}), resulting in SPARQL-queryable knowledge graphs.

  \item \textbf{Cross-domain evaluation through an anomaly detection use case.} SensorWF is evaluated on three established datasets using E1 fault injection and E2 multi-detector evaluation as concrete analytical extensions.
  \end{enumerate}


\section{Background and Related Work}
\label{sec:background}

\subsection{FAIR Workflows, Provenance, and Semantic Foundations}

The FAIR principles (Findable, Accessible, Interoperable, Reusable) mandate that computational workflows provide typed module specifications, explicitly document scientific assumptions, and include machine-readable metadata to facilitate discovery and reuse without requiring source code inspection \cite{wilkinson2016fair,garijo2022fair}. WorkflowHub operationalizes these principles through RO-Crate packaging and adoption of open standards such as CWL, Bioschemas, and the GA4GH TRS API to support workflow dissemination \cite{gustafsson2025workflowhub}.

In scientific computing, provenance, defined as the comprehensive record of inputs, transformations, software versions, and timing, is critical for verification and downstream reliability assessment. The W3C PROV Ontology (PROV-O) establishes the standard for machine-readable provenance as queryable graphs \cite{provo2013}, and is further extended by ProvONE to support workflow-specific constructs such as pipeline programs and execution states \cite{missier2013provone}.

Sensor-oriented ontologies and vocabularies, such as SSN and SOSA, provide semantic foundations for representing heterogeneous measurements, including voltage channels, ECG signals, and temperature observations, with consistent and domain-specific meaning \cite{haller2019ssn,compton2012ssn}. Knowledge graphs operationalize these models by linking engineered features to source channels, domain concepts, and analytical outputs, thereby improving both interoperability and reproducibility \cite{provo2013,missier2013provone}.

SensorWF contributes to this ecosystem in two primary ways: first, by encoding domain assumptions in a machine-readable component registry, which enables sensor-domain integration without modifying the analytical core; and second, by treating semantic annotation as an integral workflow function rather than as a post-processing step. Features are mapped to ontology classes, and relationships are exported as machine-readable graph artifacts. This design enhances FAIR compliance, particularly in terms of interoperability and reusability, by linking outputs from diverse sensor domains, such as satellite telemetry and climate observation, to explicit scientific concepts \cite{wilkinson2016fair,gustafsson2025workflowhub}. This approach builds upon previous work on assumption-aware composition \cite{vargas-acosta_automating_2022,vargas_acosta_automated_2024,vargas-acosta_towards_2025} by shifting the focus from composing non-machine-learning scientific models to constructing reusable machine learning-enabled workflows.

\subsection{Domain Adapter Patterns in Scientific Computing}

The separation of domain-specific components from reusable algorithmic cores is a foundational principle in scientific computing. The Galaxy platform \cite{afgan2018galaxy} implements this principle using standalone Tool XML wrappers. Each domain tool specifies its inputs, outputs, and execution commands, while the scheduler, data management layer, and workflow engine remain unchanged. This design confines the integration effort for new bioinformatics tools to a single wrapper definition. However, a key limitation of Galaxy's model is the lack of runtime enforcement of typed contracts between tools, and the recording of provenance only at the workflow level rather than as first-class semantic artifacts.

Time-series toolkits such as TSFEL \cite{barandas2020tsfel} employ a complementary, configuration-driven approach. Feature extraction functions are fully decoupled from the signal source, and a JSON configuration specifies which feature families are computed and at what window size. This design separates the specification of computation from its implementation. However, TSFEL lacks abstractions for data ingestion, quality assessment pipelines, and semantic or provenance outputs, functioning as a feature library rather than a comprehensive workflow architecture.

SensorWF integrates these two architectural patterns while addressing their respective limitations. The \texttt{DomainAdapter} abstract interface formalizes domain isolation, as seen in Galaxy, by establishing explicit architectural invariants with machine-readable typed input and output contracts. M3's configuration-driven feature families extend TSFEL's approach into a comprehensive pipeline that incorporates spectral features, semantic knowledge graph annotation, and PROV-O provenance export. Portability boundaries, such as M2's assumption of temporally regular sampling, are documented as SPARQL-queryable \texttt{sensorwf:hasAssumption} predicates instead of remaining implicit within the source code.

\subsection{Anomaly Detection Benchmarks and Methods}

In this work, we employ time series anomaly detection as a use case for SensorWF. 

Anomaly detection refers to the identification of observations that deviate significantly from expected behavior, including signal spikes, sensor drift, dropout, or correlated multi-channel deviations that may indicate equipment faults or physiological events. Two principal approaches are utilized: supervised methods, which are trained on labeled examples of known fault types, and unsupervised methods, which model normal behavior and flag deviations. In most cases, labeled fault data are scarce for scientific instruments, making unsupervised detection the predominant strategy. Techniques such as statistical thresholding (Z-score, rolling robust variants), ensemble isolation (IsolationForest \cite{liu2008isolation}), density-based local outlier detection (LOF \cite{breunig2000lof}), and reconstruction-error neural networks (Autoencoders) each address distinct aspects of anomalous behavior.

\subsubsection{Spacecraft Telemetry}
In spacecraft telemetry, Hundman et al. \cite{hundman2018detecting} introduced LSTM-based detection for NASA SMAP and MSL missions. The OPS-SAT benchmark \cite{ruszczak2025opssat} provides segment-level ground truth, whereas the ESA Anomaly Database (ESA-ADB) \cite{esa_adb2024} supplies event-level ground truth with affiliation-aware metrics that more accurately represent operational detection performance compared to point-wise AUC-ROC. 

\subsubsection{Biomedical Data}

In the biomedical field, the MIT-BIH Arrhythmia Database \cite{moody2001mitbih} is the standard benchmark for arrhythmia detection. Supervised deep learning approaches such as CNN-based heartbeat classification \cite{kachuee2018ecgclass} demonstrate strong cross-dataset transferability on this corpus; SensorWF targets unsupervised fault detection without reliance on rhythm labels.

\subsubsection{Climate Observation}

The Jena Climate Dataset \cite{jena_climate} is a widely used multivariate meteorological benchmark, with data recorded at 10-minute intervals from 2009 to 2016. Climate anomaly detection in this context presents challenges distinct from those associated with higher-frequency signals. Specifically, low sampling rates result in contextual anomalies, defined as deviations from expected seasonal patterns rather than absolute value thresholds, being more prevalent than point anomalies. Autoregressive methods have been employed to identify instrumental drift and calibration failures; however, these methods require stationarity assumptions that often fail during seasonal transitions \cite{martinez2020climate}. In contrast, unsupervised reconstruction-based and isolation-based approaches that operate on rolling windows are better suited to the non-stationary, multi-variable characteristics of meteorological data. Wu and Keogh \cite{wu2022benchmarks} show that single-axis difficulty scaling can produce misleading detector rankings, which motivates E1's simultaneous and independent scaling of amplitude, duration, and channel spread across all three domains.

\subsection{KISPE SATLL}

The KISPE Satellite Learning Laboratory (SATLL) provides telemetry data from multiple onboard experiments designed to demonstrate spacecraft subsystem functionality and generate sensor datasets for analysis. These experiments focus primarily on the Attitude Determination and Control System (ADCS), including accelerometer, gyroscope, and reaction wheel tests, as well as thermal monitoring experiments representative of environmental testing commonly performed on spacecraft systems. Together, these experiments generate telemetry data across multiple subsystems and sensor types, providing a useful foundation for analysis, visualization, and semantic modeling of CubeSat telemetry data.

The laboratory supports four experiment families. Three exercise the ADCS: an \emph{accelerometer} test that sweeps the CubeSat through successive orientations to record acceleration on the x, y, and z axes; a \emph{gyroscope} test that rotates the suspended satellite through a fixed sequence (90$^\circ$ counterclockwise, 180$^\circ$ clockwise, then back to origin), holding each position for several seconds; and a \emph{reaction wheel} test that actuates the wheel to rotate the satellite while the gyroscope records the resulting motion. The fourth is a longer-duration \emph{thermal} test, in which the satellite is warmed in direct sunlight, cooled indoors, and then heated by the onboard heater, producing a high-volume record across multiple independent thermal nodes. Together the four families span both stable regimes (accelerometer, thermal) and oscillatory ones (reaction wheel, gyroscope) at approximately 1\,Hz, which makes them a useful stress test for a domain-agnostic analytical core.



\section{Methodology}

Within this section the framework is described in three segments. The design principles below state the architectural invariants that make a single core reusable across domains. The component registry and module descriptions then specify what each module consumes and produces, and the domain application subsections show how those specifications are instantiated for satellite, ECG, and climate data.

\subsection{Design Principles}

The SensorWF framework is based upon five core design principles.

\textbf{Core/extension separation.} The framework differentiates a reusable core (M1 to M5) from optional use-case extensions. The core uniformly manages ingestion, quality assessment, feature extraction, semantic annotation, and provenance across all domains. Extensions, such as fault injection (E1) and anomaly detection (E2), interface with core typed outputs without modifying any core module. This separation is explicitly specified in \texttt{components.json} using an \texttt{extension} flag.

\textbf{Adapter isolation.} All domain-specific logic is encapsulated within M1 via the \texttt{DomainAdapter} abstract interface. Each adapter implements four methods: \texttt{load()}, \texttt{get\_quality\_config()}, \texttt{get\_feature\_config()}, and \texttt{get\_ontology\_path()}, thereby providing comprehensive configuration for all core modules. Integrating a new domain requires implementing only these four abstract methods in a new M1 adapter class (approximately 150--200 lines for the \texttt{DomainAdapter} interface); M2 through M5 require no modification. Use-case extensions may additionally invoke \texttt{get\_fault\_types()} as required.

\textbf{Explicit I/O contracts.} Each module defines typed input and output ports that adhere to the ProvONE \texttt{provone:Program} pattern \cite{missier2013provone}. These contracts are documented in \texttt{components.json} to facilitate programmatic discovery and composition. 


\textbf{Scientific assumption transparency.} Domain-specific assumptions, including sampling rates, quality thresholds, and window sizes, are recorded as \texttt{sensorwf:hasAssumption} predicates in both the static workflow specification and the runtime provenance trace. This method ensures that assumptions are directly SPARQL-queryable.

\textbf{Reproducibility by design.} All sources of randomness originate from a single seeded generator chain (\texttt{seed = 42}). Repeated execution of any domain entry point produces bit-for-bit identical clean data, quality reports, feature matrices, and knowledge graph artifacts. Result caching, indexed by sentinel files (\texttt{run\_summary.json}, \texttt{ml\_results.csv}), prevents redundant recomputation unless otherwise specified.

\subsection{Component Registry and Module Descriptions}

\textbf{M1: Data Ingestion.} Three concrete adapters are provided: \texttt{SatelliteAdapter} (SCOTTI v2 hex-encoded telemetry at $\sim$1\,Hz); \texttt{ECGAdapter} (MIT-BIH CSV at 360\,Hz, decimated 7$\times$ to $\approx$51\,Hz by uniform subsampling, 5-minute sessions yielding 15,000 samples); and \texttt{ClimateAdapter} (Jena CSV at 10-minute resolution, 14 channels, half-year slices). All three return a standardized DataFrame with mandatory columns \texttt{timestamp}, \texttt{elapsed\_s}, and one or more signal channels.

\textbf{M2: Quality Assessment.} Domain-agnostic quality checks are configured via the adapter's \texttt{get\_quality\_config()} dictionary: NaN rates, stuck-sensor detection ($\le k$ unique values in a window), timing regularity ($> n\times$ expected interval), per-channel Z-score flags, and optional linear-trend detection. Outputs a quality report JSON and a cleaned DataFrame.

\textbf{M3: Feature Engineering.} Nine feature families are generated per channel: (i) raw channel value, (ii) first-order difference, (iii) rolling mean, (iv) rolling standard deviation, (v) rolling skewness, (vi) rolling excess kurtosis, (vii) zero-crossing rate relative to the channel mean, (viii) spectral entropy (Shannon entropy of the rolling power spectrum computed via vectorized batched FFT), and (ix) dominant frequency in Hz (peak non-DC frequency). Rolling skewness and kurtosis are calculated using a Cython-compiled O(n) sliding-window accumulator (\texttt{scripts/\_core\_cy.pyx}, \texttt{-O3 -ffast-math}). Spectral features are computed using a three-dimensional stride-tricks FFT across all channels, and lastly, a sample-interval timing feature (\texttt{dt\_sample}) is appended. The resulting feature matrix is $N \times D$, where $D$ depends on channel count and window size.

\textbf{M4: Semantic Annotation.} Feature importances are mapped to a domain-specific OWL ontology using lexical prefix matching on feature names. The resulting knowledge graph includes OWL class-instance nodes, feature nodes, and evidence edges with the \texttt{if:featureImportance} and \texttt{if:evidenceForClass} predicates, enabling SPARQL-queryable subsystem attribution. When executed after E2, anomaly tag nodes are incorporated; in core-only mode, edges represent feature-to-class evidence derived directly from the M3 feature matrix. Outputs include RDF/Turtle files and CSV node and edge lists.

\textbf{M5: Provenance Export.} The \texttt{ProvenanceRecorder} accumulates one \texttt{prov:Activity} per module call, capturing wall-clock timestamps, input row counts, output file paths, and parameter values. SHA-256 checksums are automatically computed for all file-path entities and stored as \texttt{telwf:sha256} triples, enabling bit-level reproducibility verification. The serialized PROV-O/ProvONE Turtle document is produced as a primary workflow output, regardless of whether use-case extensions are executed.

\textbf{E1: Fault Injection (use-case only).} The procedure reads \texttt{get\_fault\_types()} and applies each morphology across three difficulty tiers, scaling amplitude, duration, and channel spread simultaneously. This multi-axis approach mitigates confounds associated with single-axis separability \cite{wu2022benchmarks}. With two variants per (type, tier) pair, E1 generates $N_{\text{faults}} \times 3 \times 2$ labeled sessions per domain.

\textbf{E2: Anomaly Detection (use-case only).} Five machine learning detectors are trained on the first 60\% of the clean session, which is temporally prior to any injected segment: (i) Z-Score with MAD-robust fusion (Iglewicz-Hoaglin threshold \cite{iglewicz1993outliers}); (ii) RobustRollingZScore with CUSUM persistence \cite{page1954cusum}; (iii) IsolationForest with a two-member random-rotation ensemble \cite{liu2008isolation}; (iv) a multi-scale MLP Autoencoder with Gaussian denoising ($\sigma\!=\!0.06$), dual sequence windows (lengths 8 and 16), and a five-layer symmetric encoder-decoder (widths $\lfloor n/3\rfloor$, $\lfloor n/8\rfloor$, $\lfloor n/16\rfloor$, $\lfloor n/8\rfloor$, $\lfloor n/3\rfloor$ for flattened window dimension $n$, tanh activations) \cite{sakurada2014autoencoder}; and (v) Local Outlier Factor (LOF) with novelty detection and PCA pre-processing (95\% variance retained) \cite{breunig2000lof}. Ensemble and density-based detectors (IF, LOF, PCA) use scikit-learn \cite{pedregosa2011sklearn}. For density-based detectors (IF and LOF), a Peaks-over-Threshold (POT) extreme-value theory calibration \cite{siffer2017evt} fits a Generalized Pareto Distribution to the tails of the training-set scores, whereas statistical detectors use the 99th percentile threshold. Per-fault and aggregate metrics, including AUC-ROC, AUC-PR, F1, FPR, recall, and the binary \texttt{event\_detected} metric, are reported by tier.

\subsection{Domain Applications}

Section~\ref{sec:background} provided an overview of each domain at the levels of physical experimentation and dataset description. The following subsections examine the specifics of the M1 adapter, present fault taxonomies, and detail module-level parameter selections that enable the operationalization of each domain within SensorWF.


\subsubsection{Satellite Telemetry (KISPE SATLL)}

The KISPE Satellite Learning Laboratory (SATLL) provides authentic telemetry (real-time measurement data transmitted from remote sensors) from four experiment families: AccelerometerTest, GyroTest, ReactionWheelTest, and ThermalTest \cite{kispe_satll}. Each session generates CDH (Command and Data Handling) data, encompassing 94 columns including measurements of thermal, power, and timing, as well as ADCS (Attitude Determination and Control System) data, covering 32 columns that include sensors for inertia measurement units (IMUs), reaction wheels, and sun sensors, at approximately 1 Hz (one measurement per second). E1 introduces 18 fault types: 16 single-channel faults (in which only one measurement channel is affected) and 2 compound faults involving multiple channels. CDH faults encompass issues such as power-rail drift (gradual voltage changes), thermal ramp (continuous temperature change), and packet dropout (missing data segments), while ADCS faults include gyro clipping (sensor saturation), magnetometer inversion (incorrect field polarity), and wheel runaway or stiction (uncontrolled or stuck motion). For each experiment family, there are 108 labeled sessions ($18 \times 3 \times 2 = 108$).

The satellite OWL ontology is generated by the pipeline at runtime. The script \texttt{satellite\_ontology.py} examines the CDH and ADCS channel lists identified during M1 and produces an OWL/XML document containing subsystem classes (\texttt{sat:ADCS}, \texttt{sat:OBDH}, \texttt{sat:EPS}, \texttt{sat:ThermalControlSubsystem}), sensor-type subclasses, and individual channels. Alignment with SSN and SOSA is achieved through the use of \texttt{sosa:Platform} and \texttt{ssn:Sensor} superclasses. A single-session M4 execution yields 1,104 knowledge graph nodes and 846 edges.

\subsubsection{Biomedical ECG (MIT-BIH Arrhythmia Database)}

The MIT-BIH Arrhythmia Database \cite{moody2001mitbih} comprises 48 thirty-minute, two-lead ambulatory ECG recordings at 360 Hz from PhysioNet \cite{goldberger2000physiobank}, each annotated beat-by-beat by cardiologists. Twenty records are selected to represent all major rhythm classes: normal sinus rhythm (100, 101, 103, 112, 113, 115), bundle branch block (106, 108, 109, 111, 118), frequent premature ventricular contractions (PVCs) and bigeminy (105, 119, 200, 205, 215), and atrial fibrillation with complex ventricular rhythms (201, 208, 213, 221). The \texttt{ECGAdapter} applies uniform subsampling by a factor of 7 ($360\,\text{Hz}\to\!\approx\!51\,\text{Hz}$) with no anti-aliasing filter. Six ECG-specific fault morphologies are introduced: baseline wander, electrode dropout, EMG burst, powerline noise (10 Hz, representing the alias of 60 Hz mains interference after decimation: $60 \bmod 50 = 10$ Hz), amplitude scaling, and lead inversion. For each record, $6 \times 3 \times 2 = 36$ labeled sessions are generated.

\subsubsection{Atmospheric Climate (Jena Climate Dataset)}

The Jena Climate Dataset \cite{jena_climate} records 14 meteorological variables, including temperature, pressure, humidity, and wind speed or direction, at 10-minute intervals from 2009 to 2016 (420,551 rows). The \texttt{ClimateAdapter} extracts half-year segments (approximately 26,000 rows each) and removes $-9999$ sentinel values. M3 employs a 24-sample rolling window (4 hours). Seven half-year sessions from 2009 to 2015 are processed, collectively covering cold and warm seasons, distinct instrument drift periods, and seven years of inter-annual variability. Six climate-specific fault morphologies are introduced: sensor drift, stuck sensor, spike burst, sensor dropout, scale error, and sign flip. For each session, $6 \times 3 \times 2 = 36$ labeled sessions are produced.


\section{Domain Application Results}

In this section we describe our empirical evaluation of SensorWF in addition to our results. Section~\ref{sec:protocol} fixes the train/test discipline shared by all three domains. The three domain subsections then report detection performance and explain each ranking in terms of session length, channel count, and fault morphology. The final two subsections cover outputs that are produced regardless of whether the detection extension runs at all, those being the M4 knowledge graph and the runtime cost of the core itself.

\subsection{Training and Evaluation Protocol}
\label{sec:protocol}

E2 applies a strict temporal train/test split to prevent data leakage. The initial 60\% of each clean baseline session is designated as training data. E1 evaluates only the remaining 40\% suffix; injected windows always commence at or after sample $\lfloor 0.6 \cdot N \rfloor$, which ensures no temporal overlap between the training prefix and any evaluated anomalous segment. Rolling-window features, with a maximum of 100 samples at 50 Hz, are negligible compared to the $\geq\!6{,}000$-sample clean gap, thereby preventing roll-back leakage. Evaluation metrics include sample-wise AUC-ROC, AUC-PR, F1, FPR, recall, and \texttt{event\_detected} (a binary indicator of whether the injected anomaly window was intercepted by at least one true positive). AUC-PR serves as the primary imbalance-aware metric, given the anomaly prevalence of $\approx\!20$--$40\%$ per injected session.

\begin{table}[t]
\caption{Cross-Domain Detection Performance}
\label{tab:crossdomain}
\centering
\scriptsize
\setlength{\tabcolsep}{2.5pt}
\begin{tabular}{@{}lrrccccc@{}}
\toprule
 & & & \multicolumn{5}{c}{\textbf{AUC-ROC / AUC-PR}} \\
\cmidrule(lr){4-8}
\textbf{Domain} & \textbf{Sess.} & \textbf{Faults} & \textbf{AE} & \textbf{IF} & \textbf{LOF} & \textbf{RRZS} & \textbf{ZS} \\
\midrule
Satellite & 1  & 18 & \textbf{0.772}/\textbf{0.581} & 0.564/0.298 & 0.549/0.280 & 0.597/0.359 & 0.610/0.394 \\
ECG     & 20 &  6 & 0.915/\textbf{0.837} & 0.880/0.762 & \textbf{0.918}/0.826 & 0.487/0.394 & 0.785/0.637 \\
Climate    &  7 &  6 & \textbf{0.835}/\textbf{0.725} & 0.720/0.577 & 0.737/0.609 & 0.536/0.399 & 0.681/0.552 \\
\bottomrule
\multicolumn{8}{l}{\scriptsize AE=Autoencoder, IF=IsolationForest, LOF=Local Outlier Factor,} \\
\multicolumn{8}{l}{\scriptsize RRZS=RobustRollingZScore, ZS=ZScore. Bold: best per domain.} \\
\multicolumn{8}{l}{\scriptsize ECG std: AE $\pm$0.061, LOF $\pm$0.048; Climate: AE $\pm$0.034, LOF $\pm$0.044.}
\end{tabular}
\end{table}
\begin{figure}[t]
\centering
\includegraphics[width=\columnwidth]{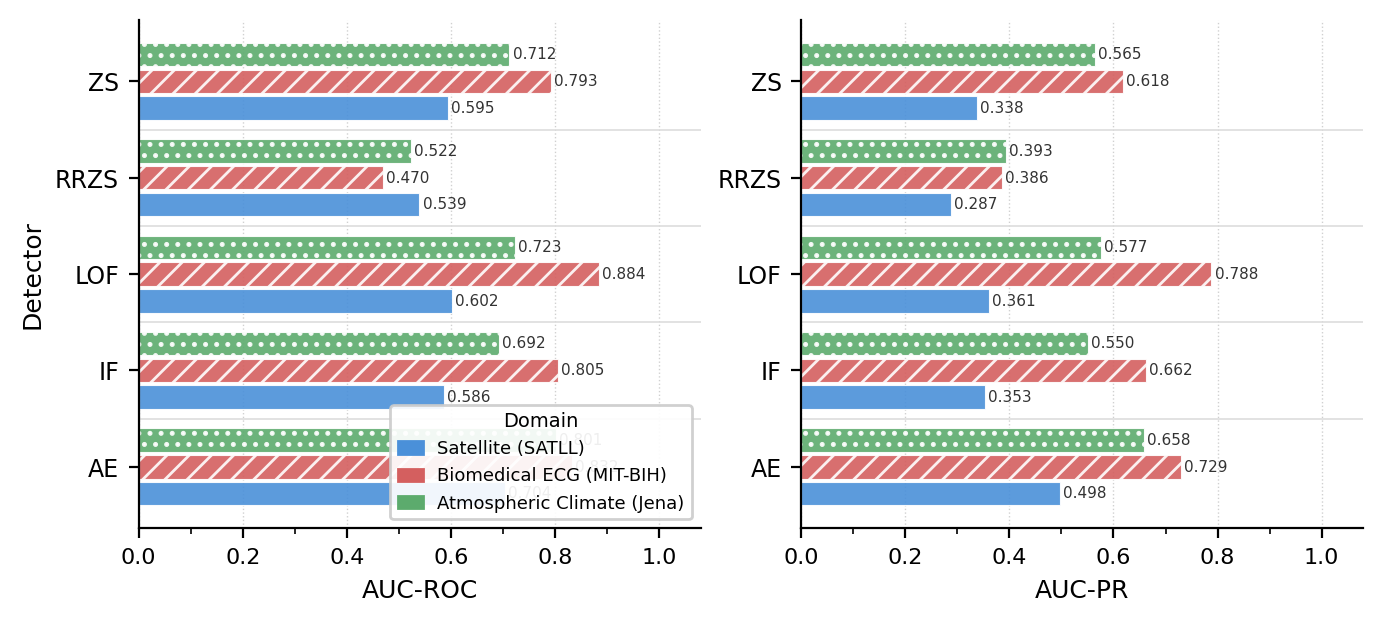}
\caption{Cross-domain detection performance for all five E2 detectors across satellite (SATLL), biomedical ECG (MIT-BIH), and atmospheric climate (Jena). Values are macro-averaged over all sessions per domain. AE=Autoencoder, IF=IsolationForest, RRZS=RobustRollingZScore, ZS=ZScore.}
\label{fig:crossdomain}
\end{figure}

\subsection{Satellite Telemetry (KISPE SATLL)}

In the satellite domain, the Autoencoder outperforms all five detectors, achieving an AUC-ROC of $0.772$ and an AUC-PR of $0.581$, which aligns with its temporal reconstruction capabilities for multi-channel CDH and ADCS fault morphologies. The detector ranking for the satellite domain differs substantially from those observed in the ECG and Climate domains: ZScore ranks second (AUC-ROC $0.610$, AUC-PR $0.394$), followed by RobustRollingZScore ($0.597$/$0.359$), while IsolationForest ($0.564$/$0.298$) and LOF ($0.549$/$0.280$) rank lowest. This reversal in the performance of density-based detectors is attributed to the short session length (333 rows), which restricts the training prefix to approximately 200 samples, insufficient for LOF and IsolationForest to construct reliable density estimates across 122 channels. The Autoencoder's reconstruction loss demonstrates reduced sensitivity to training-set size in low-row scenarios, establishing it as the most robust option for this domain. Easy-tier faults yield the highest Autoencoder AUC-ROC ($0.801$) due to larger amplitude injections, whereas Hard-tier faults result in the highest AUC-PR ($0.646$) as the longer fractional window increases the positive class rate.

\subsection{Biomedical ECG (MIT-BIH Arrhythmia Database)}

In the ECG domain, LOF achieves the highest AUC-ROC ($0.918 \pm 0.042$) and AUC-PR ($0.826 \pm 0.088$) across all 20 records, narrowly surpassing the Autoencoder (AUC-ROC $0.915 \pm 0.055$, AUC-PR $0.837 \pm 0.097$) and IsolationForest ($0.880$/$0.762$). The two leading detectors are effectively tied on AUC-ROC, while the Autoencoder marginally leads on AUC-PR ($0.837$ vs. $0.826$), indicating an advantage on harder-tier faults where longer injected windows elevate the precision-recall operating point. Six morphologically distinct fault types generate compact, well-separated clusters in the 19-feature rolling-window space, which benefits both neighborhood-density estimation and reconstruction-based approaches. Low standard deviations for IsolationForest ($\pm 0.047$) and LOF ($\pm 0.042$) indicate stable performance across the 20-record rhythm-class mix. RobustRollingZScore demonstrates poor performance (AUC-ROC $0.487$, AUC-PR $0.394$), consistent with the breakdown of its fixed-window assumption when confronted with varied ECG artifact morphologies. Per-tier results indicate strong performance across all difficulty levels for both the Autoencoder and LOF, with AUC-ROC ranging from $0.876$ (Hard) to $0.943$ (Easy) and $0.886$ (Hard) to $0.947$ (Easy), respectively. Full per-tier numeric results are provided in Appendix~\ref{app:tier}.

\subsection{Atmospheric Climate (Jena Climate Dataset)}

In the climate domain, the Autoencoder achieves the highest performance with an AUC-ROC of $0.835$ and an AUC-PR of $0.725$, exhibiting notably low variance across the seven half-year sessions ($\pm 0.023$). This consistency confirms reliable detection of sustained meteorological sensor faults over seven years of seasonal and inter-annual variability. LOF ranks second (AUC-ROC $0.737 \pm 0.041$, AUC-PR $0.609 \pm 0.055$), followed by IsolationForest ($0.720$/$0.577$). ZScore ($0.681$/$0.552$) outperforms RobustRollingZScore ($0.536$/$0.399$), with the latter's degradation attributed to its sensitivity to variable anomaly duration across tiers. Hard-tier AUC-PR values substantially exceed those of the Easy-tier for all detectors (e.g., Autoencoder: $0.856$ vs. $0.549$), driven by the larger fractional window occupied by Hard-tier injections, which increases the positive class rate. IsolationForest demonstrates the largest gap between AUC-ROC and AUC-PR ($0.720$ vs. $0.577$ overall), indicating weaker calibration at the precision-recall operating point for climate fault distributions.

\subsection{Semantic Knowledge Graph}

\begin{figure}[t]
\centering
\includegraphics[width=\columnwidth]{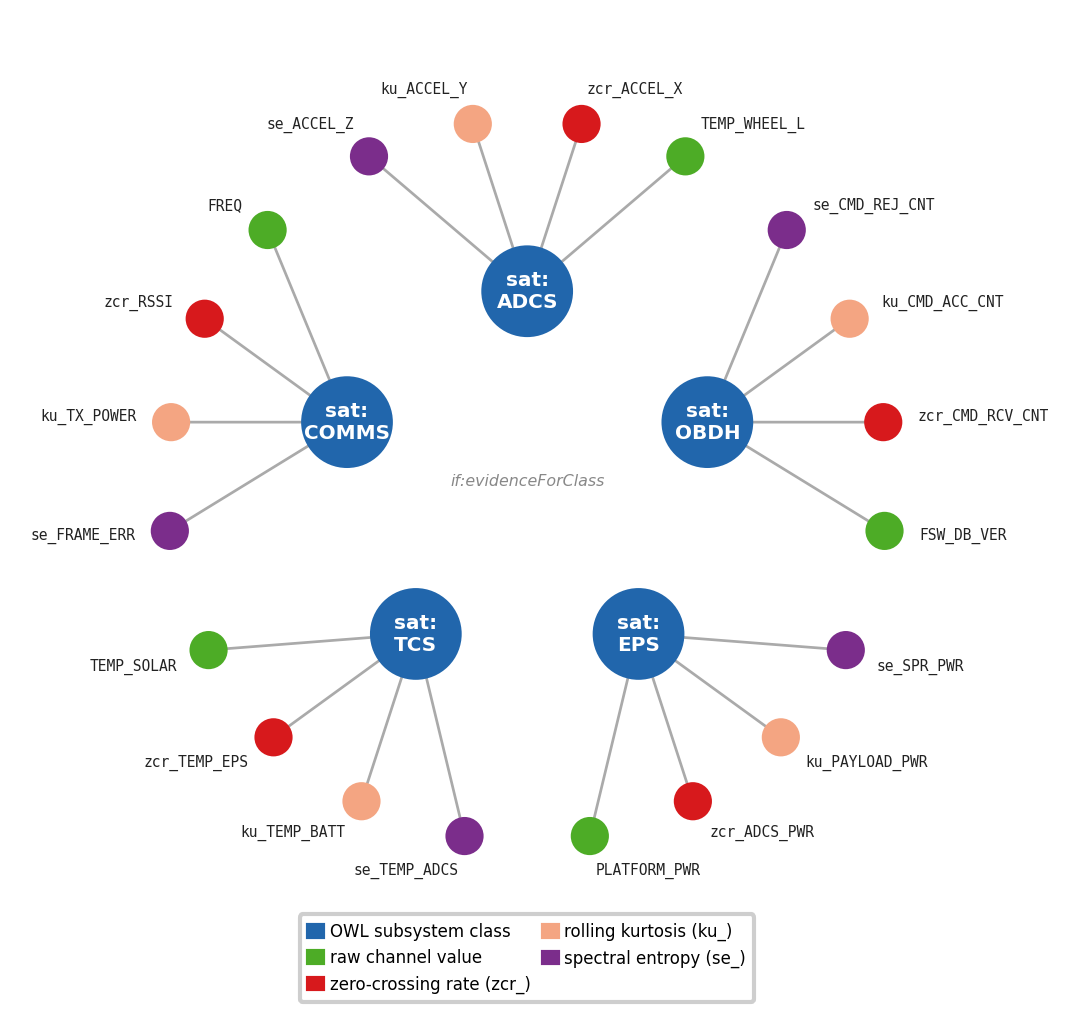}
\caption{M4 knowledge graph subgraph for the satellite domain (AccelerometerTest family). Nodes represent OWL class instances (\texttt{sat:ADCS}, \texttt{sat:OBDH}, etc.), anomaly tags, and IsolationForest feature nodes. Edges carry \texttt{if:featureImportance} and \texttt{if:evidenceForClass} predicates, enabling SPARQL queries that link detector evidence to satellite engineering subsystems without re-executing the detector.}
\label{fig:kgsubgraph}
\end{figure}

The satellite core M4 run generates a knowledge graph comprising 1,104 nodes and 846 edges. Following E2, the machine learning-enhanced knowledge graph incorporates IsolationForest feature importances, resulting in 23 ML-weighted nodes linked to subsystem classes. The most frequently queried evidence class via SPARQL is \texttt{sat:AttitudeDeterminationAndControlSubsystem}, which aligns with the ADCS-heavy channel structure. Figure~\ref{fig:kgsubgraph} presents the resulting subgraph. The \texttt{if:featureImportance} predicate enables queries of the form \texttt{SELECT ?feat ?cls WHERE \{?feat if:evidenceForClass ?cls\}} without the need to rerun the detector, thereby supporting post-hoc subsystem attribution; channel-level attribution methods \cite{condattrib2026,multiviewae2026} are identified as a complementary extension. In the ECG domain, per-record M4 runs yield compact knowledge graphs averaging 20 nodes and 9 edges each (401 nodes and 189 edges in total across 20 records), reflecting the two-channel signal structure (19 features per record). In the Climate domain, per-session M4 runs produce substantially larger graphs (987 total nodes and 882 total edges across 7 sessions) due to the 127 features derived from 14 meteorological channels.

\subsection{Runtime Characterization}
\label{sec:runtime}

Table~\ref{tab:runtime} in Appendix~\ref{app:runtime} presents per-module wall-clock times measured on a standard laptop (Apple M-series, single thread). E2 accounts for approximately 80--90\% of the total pipeline time. A detector-fit cache, keyed on detector configuration rather than fault tag, ensures that each family incurs only one training pass and re-evaluates all $N_{\text{faults}} \times 3 \times 2$ variants without redundant model refits. Addressing a previous defect that re-trained all five detectors once per fault tag (resulting in $6\times$ redundant Autoencoder fits per record) reduced ECG E2 time from over 1,000 seconds per record to the values reported. Core modules M1--M5 collectively consume less than 10\% of runtime, indicating that framework overhead is negligible compared to domain analysis costs. M3 consistently remains under 1 second across all domains due to the Cython-compiled skew/kurtosis accumulator and the batched stride-tricks FFT.


\section{Discussion}

Having established that the core transfers across three domains, this section examines what that transfer actually costs, how far the FAIR objectives were met in practice, and where the evaluation protocol falls short of operational satellite benchmarking practice.

\subsection{Generalizability and Adapter Effort}

For the three validated domains, implementing each adapter required 167 to 502 lines of Python code, whereas M2 through M5 required no code modifications. The spread is informative: the climate and ECG adapters need only 167 and 185 lines respectively, while the satellite adapter reaches 502 because it decodes hex-encoded SCOTTI frames across 122 channels and declares 18 fault morphologies. Adapter effort therefore tracks the complexity of the raw archive format rather than the complexity of the framework. The component registry enables programmatic composition: consumers query components.json for modules with a specified domain tag and receive fully specified input and output contracts, facilitating automated workflow instantiation and validation. Although only M1 is domain-specific at the code level, it is important to note that the behavior of M2 through M5 depends on adapter-supplied configuration dictionaries that encode domain knowledge, which must be provided explicitly by the adapter author. Portability limitations, such as M2's assumption of temporally regular sampling, are documented as sensorwf:hasAssumption predicates in the registry. For instance, irregular-rate or event-driven streams require time-indexed rolling windows in M3 and resampling extensions in M2.

\subsection{FAIR Assessment}

During the creation of SensorWF, aligning with the four FAIR principles was a primary objective. Reproducibility is ensured at the code level: executing any domain entry point with --seed 42 produces bit-identical CSVs, injection summaries, metrics, and knowledge graph artifacts. The only output that varies between runs is provenance.ttl, in which timestamps are intentionally different. Workflow registration on WorkflowHub with RO-Crate metadata and immutable Zenodo archives is planned for the open scientific object release.

\subsection{Relation to Satellite Anomaly Benchmarks}

The ESA-ADB benchmark \cite{esa_adb2024} recommends event-level ground truth, affiliation-aware scoring, and dynamic or extreme value theory (EVT) thresholding with postprocessing for operational satellite telemetry. OPS-SAT \cite{ruszczak2025opssat} emphasizes segment-level precision at low operating false positive rates. SensorWF partially addresses these recommendations: density-based detectors (IF, LOF) utilize peaks-over-threshold (POT) EVT calibration (Section~\ref{sec:protocol}), and an event-level event detected metric is generated for each variant. Full affiliation-aware scoring and dynamic postprocessing remain priority extensions. E2's threshold strategy and detector suite are registry-declared parameters, which supports incremental alignment with ESA-ADB protocols without requiring modifications to any core module.


\section{Limitations and Future Work}
\label{sec:limitations}

\textbf{Synthetic-only evaluation.} All ground-truth labels are generated by E1, and no real annotated anomalies are included. This limitation may overstate the generalizability of the results compared to operational benchmark performance (e.g., ESA-ADB, OPS-SAT). Evaluation using real annotated datasets is therefore a priority for future research.

\textbf{Sample-wise metrics.} AUC-ROC and AUC-PR are computed at the sample level, while the binary \texttt{event\_detected} flag provides an event-level complement for each variant. Event-wise, affiliation-aware, and range-based metrics \cite{esa_adb2024} more accurately reflect detection quality in operational contexts and are therefore priorities for future extension.

\textbf{Deep learning baselines.} The five E2 detectors do not incorporate transformer-based approaches such as iTransformer \cite{itransformer2024} or TranAD \cite{tuli2022tranad}, which require separate deep learning frameworks. The registry-based E2 architecture is compatible with these methods, and integration of at least one transformer baseline is planned as the next step.

\textbf{Assumption-aware and decision-aware ML workflows.} Future work will extend SensorWF by representing additional workflow assumptions and analytical choices as machine-readable metadata, including preprocessing choices,
feature-construction parameters, detector configurations, thresholding strategies, and evaluation
settings. This would support inspection, comparison, and
reuse of ML workflows across domains by linking selected workflow choices to downstream analytical products.



\section{Conclusion}

SensorWF demonstrates that a FAIR-annotated, generalizable workflow framework for scientific sensor time-series analysis is both feasible and practically valuable across multiple disciplines. The separation of domain-specific ingestion (M1 adapter) from a reusable analytical core (M2 to M5), along with the encoding of domain assumptions in a machine-readable component registry, allows integration and reuse of any sensor domain without modification to the analytical core. This design enables researchers to adopt, adapt, and extend a unified pipeline without the need to reimplement domain-specific components. Use-case extensions, including E1 fault injection and E2 anomaly detection, interface with core typed outputs without requiring changes to any core module, thereby demonstrating that the architecture accommodates a wide range of downstream analyses.

Cross-domain validation involving spacecraft telemetry (KISPE SATLL, one session, 18 fault types), ambulatory ECG (20 MIT-BIH records covering all major rhythm classes), and atmospheric climate data (seven Jena half-year sessions) demonstrates that a consistent detection architecture can be maintained from a single codebase, with domain-specific configuration restricted to M1. For satellite data, the Autoencoder achieves the highest performance (AUC-ROC $0.772$, AUC-PR $0.581$), whereas density-based detectors perform least effectively due to the short session, which limits the training prefix to approximately 200 samples. In ECG analysis, LOF and the Autoencoder yield comparable results (AUC-ROC $0.918\pm0.042$ and $0.915\pm0.055$, AUC-PR $0.826$ and $0.837$, respectively), with the Autoencoder narrowly surpassing in AUC-PR. For climate data, the Autoencoder again leads (AUC-ROC $0.835\pm0.023$, AUC-PR $0.725$) and demonstrates low variance across seven years of sessions. The evaluation protocol is enhanced by EVT-calibrated thresholding for density-based detectors, rolling-window sensitivity analysis, and an event-level detection metric. Key limitations, such as reliance on synthetic evaluation, use of point-wise metrics, and the absence of transformer baselines, are explicitly documented as \texttt{sensorwf:hasAssumption} predicates. KISPE SATLL data and all code will be made available on GitHub as an open scientific object.

\section{Acknowledgements}

Special thanks to Dr. Jerry Sellers for explaining the Kipse CubeSat dashboard. Special thanks to the College of Engineering at Embry-Riddle Aeronautical University for providing access to the CubeSat.
This project is supported by the Center for Aerospace Resilient Systems at Embry-Riddle Aeronautical University under an internal grant.

\bibliographystyle{IEEEtran}
\bibliography{main}

\appendices

\section{M1 Domain Adapter Implementation}
\label{app:adapter}

Listing~\ref{lst:adapter} gives an abridged \texttt{ClimateAdapter}, the shortest of the three validated M1 adapters at 167 lines. The four methods shown constitute the entire domain-specific surface of the framework; M2--M5 consume their outputs without modification.

\begin{lstlisting}[language=Python,caption={Abridged \texttt{ClimateAdapter}},label={lst:adapter}]
class ClimateAdapter(DomainAdapter):
    channels  = _CHANNELS       # 14 met. channels
    native_hz = 1.0 / 600.0     # 10-min sampling

    def load(self, path, **kw) -> pd.DataFrame:
        df = pd.read_csv(path, parse_dates=["Date Time"])
        df = df.rename(columns=_COL_MAP)
        df[wv] = df[wv].replace(-9999.0, np.nan)
        df["elapsed_s"] = (df.timestamp
                           - t0).dt.total_seconds()
        return df[["timestamp","elapsed_s"]+_CHANNELS]

    def get_quality_config(self) -> dict:
        return {"stuck_unique_max": 3,
                "zscore_threshold": 3.5,
                "expected_dt_s":  600.0,
                "gap_multiplier":   3.0}

    def get_feature_config(self) -> dict:
        return {"channels": _CHANNELS, "window": 24}

    def get_ontology_path(self) -> str:
        return "results/ontologies/climate.owl"
\end{lstlisting}


\section{Per-Module Runtime Breakdown}
\label{app:runtime}

Table~\ref{tab:runtime} disaggregates wall-clock time by module for a full satellite run (4 experiment families, 18 fault types, 3 difficulty tiers, 2 variants each) on a single laptop core. It is the basis for the $<$10\% core-overhead result reported in Section~\ref{sec:runtime}.

\begin{table}[h]
\caption{Per-Module Runtime (Satellite, 4 Families, 2 Variants, 3 Tiers)}
\label{tab:runtime}
\centering
\scriptsize
\setlength{\tabcolsep}{3.5pt}
\begin{tabular}{@{}lrrl@{}}
\toprule
\textbf{Module} & \textbf{Time (s)} & \textbf{\%} & \textbf{Notes} \\
\midrule
M1 Ingestion     & $<$1     & $<$1  & Parse SCOTTI archive \\
M2 Quality       & $<$1     & $<$1  & Clean + flag channels \\
M3 Features      & $<$1     & $<$1  & 9 families; Cython O($n$) + batched FFT \\
E1 Injection     & $\sim$5  & $\sim$1  & Generate 432 labelled CSVs \\
E2 Detection     & $\sim$400 & $\sim$90 & Train 5 ML detectors; config-keyed cache \\
M4 Semantic      & $\sim$30 & $\sim$7  & Build KG from ML evidence + OWL \\
M5 Provenance    & $<$1     & $<$1  & Serialize PROV-O trace + SHA-256 \\
\midrule
Total            & $\sim$440 & 100  & Single laptop, no GPU \\
\bottomrule
\end{tabular}
\end{table}


\section{ECG Detection Performance by Difficulty Tier}
\label{app:tier}

Table~\ref{tab:tier} disaggregates the ECG AUC-ROC and AUC-PR results (averaged across all 20 MIT-BIH records) by the three E1 difficulty tiers. AUC-PR increases monotonically from Easy to Hard for all detectors because Hard-tier injections occupy a larger fractional window, reducing effective class imbalance. AUC-ROC is highest on Easy for AE and LOF, reflecting their superior detection of the largest-amplitude faults; IF follows the same trend. RobustRollingZScore is near-random ($\approx\!0.49$) across all tiers, confirming its insensitivity to the varied ECG artifact morphologies.

\begin{table}[h]
\caption{ECG AUC-ROC / AUC-PR by Difficulty Tier (20 Records, 6 Fault Types)}
\label{tab:tier}
\centering
\scriptsize
\setlength{\tabcolsep}{2.5pt}
\begin{tabular}{@{}llrrrrr@{}}
\toprule
\textbf{Tier} & \textbf{Det.} & \textbf{AUC-ROC} & \textbf{AUC-PR} & \textbf{Recall} & \textbf{FPR} & \textbf{F1} \\
\midrule
Easy   & AE   & \textbf{0.943} & 0.762 & 0.806 & 0.072 & 0.674 \\
       & IF   & 0.906 & 0.638 & 0.552 & 0.037 & 0.517 \\
       & LOF  & \textbf{0.947} & 0.729 & 0.841 & 0.113 & 0.631 \\
       & RRZS & 0.482 & 0.153 & 0.055 & 0.028 & 0.075 \\
       & ZS   & 0.809 & 0.441 & 0.242 & 0.048 & 0.218 \\
\midrule
Medium & AE   & \textbf{0.927} & 0.861 & 0.768 & 0.087 & 0.753 \\
       & IF   & 0.889 & 0.793 & 0.496 & 0.044 & 0.514 \\
       & LOF  & \textbf{0.921} & 0.851 & 0.798 & 0.135 & 0.763 \\
       & RRZS & 0.483 & 0.399 & 0.059 & 0.028 & 0.094 \\
       & ZS   & 0.789 & 0.671 & 0.219 & 0.048 & 0.239 \\
\midrule
Hard   & AE   & 0.876 & \textbf{0.888} & 0.660 & 0.078 & 0.690 \\
       & IF   & 0.844 & 0.854 & 0.388 & 0.043 & 0.430 \\
       & LOF  & 0.886 & \textbf{0.899} & 0.701 & 0.097 & 0.742 \\
       & RRZS & 0.495 & 0.630 & 0.062 & 0.029 & 0.104 \\
       & ZS   & 0.757 & 0.799 & 0.160 & 0.043 & 0.202 \\
\bottomrule
\multicolumn{7}{l}{\scriptsize AE=Autoencoder, IF=IsolationForest, RRZS=RobustRollingZScore, ZS=ZScore.} \\
\multicolumn{7}{l}{\scriptsize Bold: best AUC-ROC and AUC-PR per tier. Averaged across 20 records, 6 fault types.}
\end{tabular}
\end{table}

\section{M4 Knowledge Graph SPARQL Query Examples}
\label{app:sparql}

The following queries illustrate the post-hoc analytical value of the M4 knowledge graph without re-executing any ML detector.

\noindent\textbf{Q1: Which satellite subsystems carry the strongest anomaly evidence?}
\begin{lstlisting}[basicstyle=\ttfamily\scriptsize]
PREFIX if: <http://sensorwf.org/interpretability#>
PREFIX sat: <http://sensorwf.org/satellite#>
SELECT ?subsystem (SUM(?imp) AS ?total_evidence)
WHERE {
  ?feat if:belongsToSubsystem ?subsystem ;
        if:featureImportance ?imp .
}
GROUP BY ?subsystem ORDER BY DESC(?total_evidence)
\end{lstlisting}

\noindent\textbf{Q2: Top-5 features by IsolationForest importance for gyro faults:}
\begin{lstlisting}[basicstyle=\ttfamily\scriptsize]
PREFIX if: <http://sensorwf.org/interpretability#>
SELECT ?feat ?cls ?imp
WHERE {
  if:tag:A2_gyro_clipping if:importantFeature ?feat .
  ?feat if:featureImportance ?imp ;
        if:mapsToSensorClass ?cls .
}
ORDER BY DESC(?imp) LIMIT 5
\end{lstlisting}

These queries execute against the RDF/Turtle graph (\texttt{if\_ontology\_graph.ttl}) produced by M4 using any SPARQL~1.1 endpoint (e.g., Apache Jena Fuseki, RDFLib). No Python or ML libraries are required for post-hoc subsystem attribution.

\section{M3 Rolling-Window Ablation -- Full Numeric Results}
\label{app:ablation}

Table~\ref{tab:ablation_full} gives the complete per-detector AUC-ROC and AUC-PR values. Results are averaged over all fault types and three difficulty tiers for 5 MIT-BIH ECG records.

\begin{table}[h]
\caption{M3 Window Ablation -- Mean AUC-ROC / AUC-PR (ECG, 5 Records)}
\label{tab:ablation_full}
\centering
\scriptsize
\setlength{\tabcolsep}{3pt}
\begin{tabular}{@{}lrrrrr@{}}
\toprule
 & \multicolumn{5}{c}{\textbf{AUC-ROC (AUC-PR)}} \\
\cmidrule{2-6}
$w$ & \textbf{AE} & \textbf{IF} & \textbf{LOF} & \textbf{RRZS} & \textbf{ZS} \\
\midrule
  15 & 0.775 (0.651) & 0.707 (0.554) & 0.854 (0.768) & 0.449 (0.370) & 0.688 (0.523) \\
  30 & 0.797 (0.671) & 0.758 (0.594) & 0.869 (0.777) & 0.452 (0.371) & 0.756 (0.575) \\
  50 & 0.793 (0.671) & 0.803 (0.647) & 0.868 (0.773) & 0.462 (0.382) & 0.794 (0.612) \\
 100 & 0.807 (0.694) & 0.816 (0.671) & 0.876 (0.780) & 0.464 (0.388) & 0.806 (0.625) \\
\bottomrule
\end{tabular}
\end{table}

\section{M5 PROV-O Runtime Provenance Trace}
\label{app:provenance}

\begin{figure}[h]
\centering
\includegraphics[width=\columnwidth]{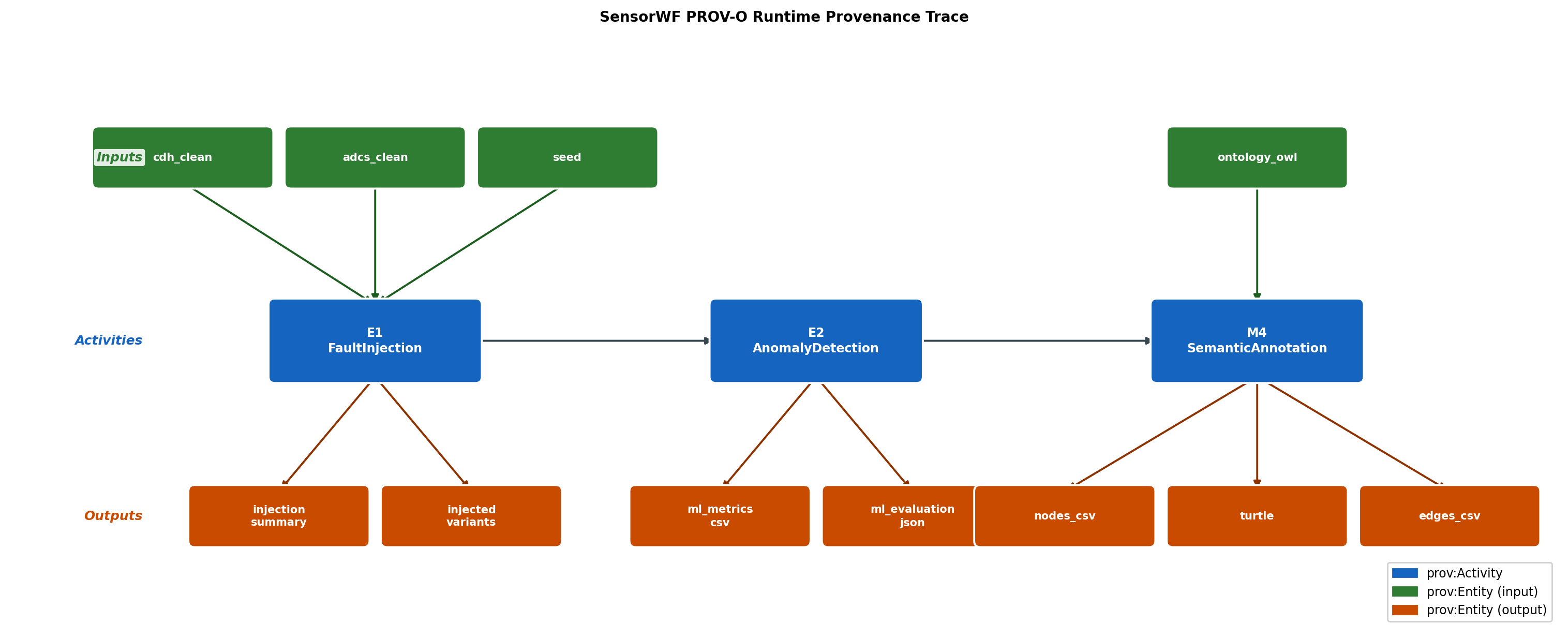}
\caption{PROV-O runtime provenance trace for the satellite anomaly detection use case, parsed from \texttt{anomaly\_provenance.ttl}. Blue boxes: \texttt{prov:Activity} nodes (E1 fault injection, E2 anomaly detection, M4 semantic annotation). Green boxes: input \texttt{prov:Entity} nodes (raw telemetry files, parameters). Orange boxes: output \texttt{prov:Entity} nodes (injected variants, ML evaluation JSON, knowledge graph CSVs). Arrows show \texttt{prov:used} (green) and \texttt{prov:wasGeneratedBy} (orange) relations. SHA-256 checksums are stored for all file-path entities.}
\label{fig:provenance}
\end{figure}

Fig.~\ref{fig:provenance} shows the PROV-O/ProvONE runtime trace produced by M5 for the satellite anomaly detection use case, parsed directly from \texttt{anomaly\_provenance.ttl}. Three \texttt{prov:Activity} nodes, E1 (fault injection), E2 (anomaly detection), and M4 (semantic annotation), are arranged in execution order with their input and output \texttt{prov:Entity} nodes. SHA-256 checksums are stored as \texttt{telwf:sha256} triples for all file-path entities. The trace is fully queryable via any SPARQL~1.1 endpoint.

\end{document}